\documentclass[aps,10pt,twocolumn,showpacs,preprintnumbers,amsmath,amssymb,prl,
superscriptaddress,floatfix]{revtex4-1}

\usepackage{graphicx}

\newcommand{\ket}[1]{\left| #1 \right\rangle}
\newcommand{\bra}[1]{\left\langle #1 \right|}
\newcommand{\eff}{\text{eff}}
\newcommand{\tint}{\text{int}}
\newcommand{\hc}{\text{h.c.}}

\begin{document}

\title{Inelastic electron backscattering in a generic helical edge channel}

\author{Thomas~L.~Schmidt}\affiliation{Department of Physics, Yale University, 217 Prospect Street, New Haven, CT 06520, USA}

\author{Stephan Rachel}\affiliation{Department of Physics, Yale University, 217 Prospect Street, New Haven, CT 06520, USA}

\author{Felix von Oppen}\affiliation{\mbox{Dahlem Center for Complex Quantum Systems and Fachbereich Physik, Freie Universit\"at Berlin, 14195 Berlin, Germany}}

\author{Leonid~I.~Glazman}\affiliation{Department of Physics, Yale University, 217 Prospect Street, New Haven, CT 06520, USA}

\date{\today}

\begin{abstract}
We evaluate the low-temperature conductance of a weakly interacting one-dimensional helical liquid without axial spin symmetry. The lack of that symmetry allows for inelastic backscattering of a single electron, accompanied by forward-scattering of another. This joint effect of weak interactions and potential scattering off impurities results in a temperature-dependent deviation from the quantized conductance, $\delta G \propto T^4$. In addition, $\delta G$ is sensitive to the position of the Fermi level. We determine numerically the parameters entering our generic model for the Bernevig-Hughes-Zhang Hamiltonian of a HgTe/CdTe quantum well in the presence of Rashba spin-orbit coupling.
\end{abstract}


\maketitle

A key feature of a 2D topological insulator (TI) is the presence of gapless edge states at its boundaries with a ``normal'' insulator or the vacuum \cite{qi11,hasan10}. If the system is time-reversal (TR) invariant, the counter-propagating states of the same energy (carrying momenta $k$ and $-k$) form a Kramers doublet, which makes elastic backscattering off a potential scatterer impossible~\cite{kane05b,bernevig06b}. Thus, potential scatterers on their own cannot prevent electrons from ballistic propagation along helical edge states. The result is a quantized, temperature-independent universal conductance of $G_0 = e^2/h$ per helical edge.

On the other hand, inelastic scattering due to a combination of electron-electron interactions and a potential which violates translational invariance along the helical edge may affect its conductance; weak interactions lead to a temperature-dependent correction reducing the conductance compared to its universal value. The existing theories~\cite{kane05b,bernevig06b,xu06,wu06,budich11} predict a power-law temperature dependence (unlike in the quantum Hall effect \cite{wei85}) and apply to interacting helical edges with conserved $S_z$ component of the electron spin. In the presence of such an auxiliary symmetry, the lowest-order processes affecting the conductance involve backscattering of electron pairs. Such two-particle backscattering may result from the presence of a lattice potential (Umklapp process) or from an inhomogeneity violating the translational invariance of the helical edge. In the former case, the temperature dependence of the leading correction to $G_0$ is $\delta G\propto T^5$, if the Fermi level is tuned to the TR invariant point of the electron spectrum \cite{kane05}. In the latter case, the two-particle backscattering off an impurity results in $\delta G\propto T^6$ \cite{wu06}, as follows from a straightforward phase space argument. If the Fermi level is shifted away from the TR invariant point, the Umklapp processes require activation energy, and are therefore exponentially suppressed at low temperatures. For the two-particle backscattering off impurities, one would expect only a weak sensitivity of $\delta G$ to the position of the Fermi level.

The prediction of a 2D topological insulator state in HgTe/CdTe quantum well heterostructures~\cite{bernevig06} prompted experiments which indeed found a low-temperature conductance close to $G_0$ for structures with the proper quantum-well thickness \cite{koenig07,koenig08}. The minimal model~\cite{bernevig06} of Bernevig, Hughes, and Zhang (BHZ) is a block-diagonal $4\times 4$ matrix Hamiltonian acting in the space of four bands originating from two spins and two orbital states. This minimal model assumes axial and inversion symmetry around the growth axis of the heterostructure ($z$ axis), which carries over to the edge states obtained within the BHZ model. The electron spins in these helical edge eigenstates are indeed oriented along the $\pm z$ direction at any momentum $k$.

The assumed axial symmetry of a HgTe/CdTe heterostructure, even if it exists in the original band-structure model, may be lifted by a gate-induced electric field in the $z$ direction. This results in a $k$-dependent Rashba spin-orbit interaction (SOI)~\cite{rothe10}, and the $S_z$ component of the electron spin is generally no longer conserved \cite{virtanen12}. We expect the absence of $S_z$ symmetry to be a rather generic property of helical edge states, which can also be realized in other models.

In this work, we evaluate the correction $\delta G$ to the universal conductance as a function of temperature and Fermi momentum (measured from the TR invariant point, $k = 0$) for a generic helical edge. We will show that if the temperature is low and the Fermi momentum is away from $k=0$, $\delta G$ is dominated by the combined effects of interaction and potential scattering off the disorder potential. In the absence of axial symmetry we find $\delta G\propto T^4$, which is stronger than the aforementioned $\delta G\propto T^6$ in the $S_z$-symmetric case. In addition, $\delta G$ acquires a substantial dependence on the Fermi level, increasing with its detuning from the TR invariant point, see Eq.~(\ref{eq:deltaGintV}) below. Moreover, inelastic backscattering of a single electron (with energy transfer to another particle-hole pair) is possible even without involvement of disorder when one of the participating states is at $k=0$. Similar to the two-particle Umklapp process in the axially-symmetric model~\cite{kane05}, these processes lead to $\delta G\propto T^5$ if the Fermi momentum is tuned to the TR invariant point, and to a thermal activation law for $\delta G(T)$ if the Fermi momentum is tuned away from $k=0$, see Eqs.~(\ref{eq:deltaGint1}) and (\ref{eq:deltaGint2}). The crossover between the inelastic backscattering involving the $k=0$ state and processes utilizing the disorder potential may lead to a non-monotonic dependence of $\delta G$ on the Fermi momentum at fixed temperature. The magnitude of the correction to the universal conductance and the details of the crossover depend on the specific interaction and disorder potentials. However, the very existence of the processes we consider rests on the rotation of the spin quantization axis with $k$ for an ideal free-electron helical edge. We determine that rotation explicitly by solving numerically the Kane-Mele~\cite{kane05} and BHZ~\cite{bernevig06} models with added Rashba SOI. The exponents of the $T$-dependence of $\delta G$ in Eqs.~(\ref{eq:deltaGintV}) and (\ref{eq:deltaGint1}) result from phase space constraints on the scattering events, together with the dependence of the scattering amplitudes on the electronic momenta, see Eqs.~(\ref{Heff1}) and (\ref{Heff2}).

The eigenstates of a translation invariant 1D helical system can be labeled by their momenta $k$. If we assume that the system is TR invariant, Kramers theorem ensures that for any $k$ there exist two degenerate orthogonal eigenstates, created by the operators $\psi^\dag_{+,k}$ and $\psi^\dag_{-,-k}$, which are related by the TR operator $\Theta$, e.g., $\Theta \psi_{\pm,k} \Theta^{-1} = \pm \psi_{\mp,-k}$. The kinetic energy has the general form,
\begin{align}\label{eq:H0gen}
    H_0 = \sum_{k} \left[ \epsilon(k) \psi^\dag_{+,k} \psi_{+,k} + \epsilon(-k) \psi^\dag_{-,k} \psi_{-,k} \right].
\end{align}
For momenta close to the Fermi momentum, $k \approx k_F$, the operators $\psi^\dag_{+,k}$ and $\psi^\dag_{-,-k}$ create right-moving and left-moving electrons, respectively, propagating with velocities $\pm v_F$, where $v_F = d\epsilon(k)/dk|_{k=k_F}$.

In a generic helical liquid, the electron spin component along a fixed $z$ direction does not have to be a good quantum number. The field operators $\psi_{\sigma,k}$ of an electron with momentum $k$ and spin projection $\sigma = \uparrow,\downarrow$ along the $z$-axis are related to the operators $\psi_{\pm,k}$ by a momentum-dependent $SU(2)$ matrix $B_k$,
\begin{align}\label{eq:Bk}
 \begin{pmatrix}
      \psi_{\uparrow,k} \\
      \psi_{\downarrow,k} \\
    \end{pmatrix}
=
    B_k
    \begin{pmatrix}
      \psi_{+,k} \\
      \psi_{-,k} \\
    \end{pmatrix}.
\end{align}
The normalization and the orthogonality of the momentum eigenstates is reflected in the unitarity condition $B_k^\dag B_k = \text{diag}(1,1)$. Moreover, TR invariance entails the symmetry $B_k=B_{-k}$ which follows by comparing how TR affects $\psi_\sigma$ and $\psi_\alpha$ ($\alpha = \pm$). This also ensures that states created by $\psi^\dag_{\alpha,k}$ and $\psi^\dag_{-\alpha,-k}$ at opposite momenta always have opposite spins, e.g., $[B^\dag_k B_{-k}]^{+-} = 0$. As one consequence, elastic backscattering between such states is prohibited for nonmagnetic impurities.

We take the electron-electron interaction to depend only on
the distance between the electrons so that $H_{\tint} = \int dx dx' U(x-x') \rho(x) \rho(x')$, where $\rho(x) = \rho_\uparrow(x) + \rho_\downarrow(x)$ is the total particle density. When expressed in terms of the eigenstates of $H_0$, this becomes
\begin{align}\label{eq:Hint}
    H_{\tint} &= \frac{1}{L}  \sum_{kk'q} \sum_{\alpha\beta\alpha'\beta'=\pm} U(q) [B^\dag_k B_{k-q}]^{\alpha\beta} [B^\dag_{k'} B_{k'+q}]^{\alpha'\beta'} \notag \\
&\times \psi^\dag_{\alpha,k}\psi_{\beta,k-q} \psi^\dag_{\alpha',k'} \psi_{\beta',k'+q},
\end{align}
where $L$ denotes the length of the helical edge and $U(q)$ is the Fourier transform of $U(x)$.

When considering an impurity violating translational invariance along the edge, we concentrate on local perturbations $V(x)$ interacting with the electron density, $H_V = \int dx V(x) \rho(x)$. Expressed in terms of $\psi_{\pm,k}$,
\begin{align}\label{eq:HV}
 H_V
&=
 \frac{1}{L} \sum_{k_1k_2}  \sum_{\alpha\beta= \pm} V(k_1 - k_2) [ B^\dag_{k_1} B_{k_2} ]^{\alpha\beta}
  \psi^\dag_{\alpha,k_1} \psi_{\beta,k_2},
\end{align}
where $V(k)$ is the Fourier transform of $V(x)$. The total Hamiltonian $H = H_0 + H_{int} + H_V$ is TR invariant, as can be seen explicitly using the unitarity and $k \to -k$ symmetry of $B_k$.

As we are interested in the conductance of edge states, we concentrate on temperatures low compared to the bulk gap and linearize the single-particle spectrum around the Fermi momentum, $\epsilon(k) = v_F ( k - k_F)$. We also make some simplifying assumptions about the form of $B_k$. A $k$-independent $B_k$ describes a constant rotation of the spin quantization axis. In this case, the spins of right- and left-movers are still opposite, irrespective of momentum. Due to the symmetry $B_k = B_{-k}$ and unitarity, the leading terms in $B_k$ for small momenta $k\ll k_0$ can be written as
\begin{align}\label{eq:Bk_approx}
 B_k = \begin{pmatrix}
        1-k^4/(2k_0^4) & - k^2/k_0^2 \\
        k^2/k_0^2 & 1-k^4/(2k_0^4)
       \end{pmatrix}.
\end{align}
Here, $k_0$ parametrizes the scale on which the spin quantization axis rotates with $k$. As confirmed for certain microscopic models below, Eq.~(\ref{eq:Bk_approx}) represents the generic form of $B_k$ up to order $(k/k_0)^2$. Higher-order terms in $k$ only give rise to subleading corrections to the conductance. Finally, we neglect the momentum-dependence of the interaction and scattering potentials, and assume $U(q) = U_0$ and $V(k) = V_0$.

Combining Eqs.~(\ref{eq:Hint}) and (\ref{eq:Bk_approx}), we find a TR invariant interaction Hamiltonian with the structure
\begin{align}
H_{\tint} &\propto \sum_{kk'q} \frac{(k^2 - k'^2)U_0}{k_0^2}
\psi^\dag_{+,k+q} \psi^\dag_{-,k'-q} \psi_{+,k'} \psi_{+,k} \notag\\
&- (\psi_+ \leftrightarrow \psi_-) + \hc \label{Heff1}
\end{align}
$H_{\tint}$ describes backscattering of a single particle, accompanied by the creation of a comoving particle-hole pair. The conductance correction is proportional to the rate of this process, which can be calculated using Fermi's golden rule. Of the three integrals over the momenta in the final state (two particles, one hole), two are canceled by energy and momentum conservation. At low $T$ and $k_F = 0$, the remaining momentum is of order $k\sim T/v_F$. The $\propto k^2$ scaling of $H_{\tint}$ then yields $\delta G_{\rm int}\propto T^5$. Interactions of the form (\ref{eq:Hint}) also cause two-particle backscattering processes. However, these result in a contribution $\delta G \propto T^9$, and are thus subleading with respect to the two-particle backscattering amplitude considered in Ref.~\cite{kane05}.

The presence of impurities relaxes the requirement of momentum conservation in the scattering process. The corresponding low-energy effective Hamiltonian, applicable for $k_F\neq 0$ and $T\ll v_F|k_F|$, and derived within perturbation theory in $H_{\tint}$ and $H_V$ has the structure
\begin{align}
H_{V,\tint}^{\eff}&\propto  \!\!\sum_{kk'qq'} \!\! \frac{(k - k')k_F U_0 V_0}{k_0^2v_F}
 \psi^\dag_{+,k+q+q'} \psi^\dag_{-,k'-q} \psi_{+,k'} \psi_{+,k}\notag\\
&+ (\psi_+ \leftrightarrow \psi_-) + \hc \label{Heff2}
\end{align}
The scattering rate now involves an additional momentum integration. Combined with the $\propto k$ scaling of $H^{\eff}_{V,\tint}$, one finds $\delta G_{V,\tint}\propto T^4$.

For the detailed evaluation of the conductance, the left and right ends (at $x=\mp L/2$) of the helical edge are coupled to electron reservoirs which are held at the same temperature $T$, but at slightly different chemical potentials $\mu_L = V/2$ and $\mu_R = -V/2$, respectively. In the clean, noninteracting limit, the conductance $G_0 = e^2/h$ is temperature-independent.

\begin{figure}[t]
  \centering
  \includegraphics[width=\columnwidth]{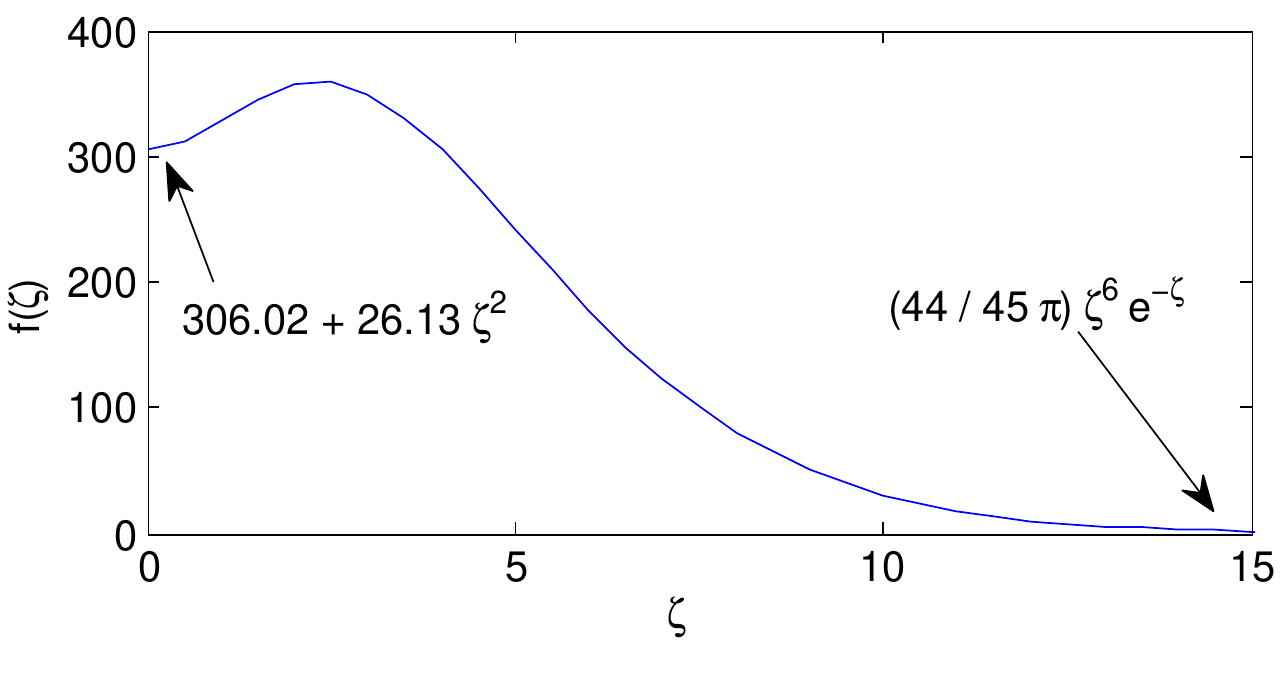}
\caption{Dimensionless factor $f(\zeta) = f(v_F k_F/T)$ describing the dependence of the interaction-induced conductance correction on the Fermi momentum, see Eq.~(\ref{eq:Gint}).}
  \label{fig:InteractionBS}
\end{figure}

We calculate the change in conductance $\delta G$ due to interactions and impurity scattering using perturbation theory in $H_{\tint}$ and $H_V$. The inelastic backscattering current is determined by the transition rate $2\pi |\bra{f} \hat{T} \ket{i}|^2 \delta(\epsilon_f - \epsilon_i)$ between initial states $\ket{i} = \psi^\dag_{\alpha_{i1},k_{i1}} \psi^\dag_{\alpha_{i2},k_{i2}} \ket{0}$ (with energy $\epsilon_i$) and final states $\ket{f} = \psi^\dag_{\alpha_{f1},k_{f1}} \psi^\dag_{\alpha_{f2},k_{f2}} \ket{0}$ (with energy $\epsilon_f$), weighted by thermal occupation factors. For single-particle backscattering $\alpha_{i1} \alpha_{i2} \alpha_{f1} \alpha_{f2} = -1$. The $\hat{T}$-matrix satisfies the equation \cite{ziman_book}
\begin{align}\label{tmatrix}
 \hat{T} = (H_{\tint} + H_{V}) + (H_{\tint} + H_V) \frac{1}{\epsilon_i - H_0} \hat{T}.
\end{align}
In the absence of interactions, $U_0 = 0$, backscattering is forbidden by TR invariance. The first-order term in the interaction, $\hat{T}=H_\tint$, yields
\begin{align}
\delta G_{\tint} &= \frac{e^2}{h}Lk_0\left(\frac{U_0}{v_F}\right)^2\left(\frac{T}{v_Fk_0}\right)^5 f\left(\frac{v_F|k_F|}{T}\right) , \notag\\
f(\zeta) &= \frac{8}{\pi}\!\int_{-\infty}^\infty dx_{1} dx_{2} \left( x_{1}^2 - x_{2}^2  \right)^2
 n_F(x_1 - \zeta) n_F(x_2 - \zeta) \notag\\
&\times [ 1 - n_F(x_1 + x_2 - \zeta)][1 - n_F(-\zeta)]\,. \label{eq:Gint}
\end{align}
Here, $n_F(x) = 1/(e^{x} + 1)$ is the Fermi function, and the function $f(\zeta)$, plotted in Fig.~\ref{fig:InteractionBS}, describes the dependence on the Fermi energy. The non-monotonic behavior of $f(\zeta)$ is a consequence of the $k$-dependence of the matrix elements of the interaction Hamiltonian, see Eq.~(\ref{Heff1}). With the asymptotes $f(\zeta = 0)\approx 306.02$ and $f(\zeta)=(44/45\pi)\zeta^{6}\exp(-\zeta)$ for $\zeta\gg 1$, we obtain
\begin{align}
\delta G_{\tint}\approx 306.02 \frac{e^2}{h}Lk_0\left(\frac{U_0}{v_F}\right)^2\left(\frac{T}{v_Fk_0}\right)^5
\label{eq:deltaGint1}
\end{align}
for $k_F = 0$, and
\begin{align}
\delta G_{\tint} \approx \frac{44}{45 \pi} \frac{e^2}{h}Lk_0\!\left(\frac{U_0}{v_F}\right)^2\!\left(\frac{k_F}{k_0}\right)^6\!
\frac{v_F k_0}{T}e^{-\frac{v_F|k_F|}{T}}
\label{eq:deltaGint2}
\end{align}
for $v_F|k_F|\gg T$. The correction $\delta G_{\tint}$ in Eq.~(\ref{eq:deltaGint2}) is activated because energy and momentum conservation require that the counter-propagating particle in the final state be created at zero momentum, which is deep within the Fermi sea.

\begin{figure}[t]
  \centering
  \includegraphics[width=5cm]{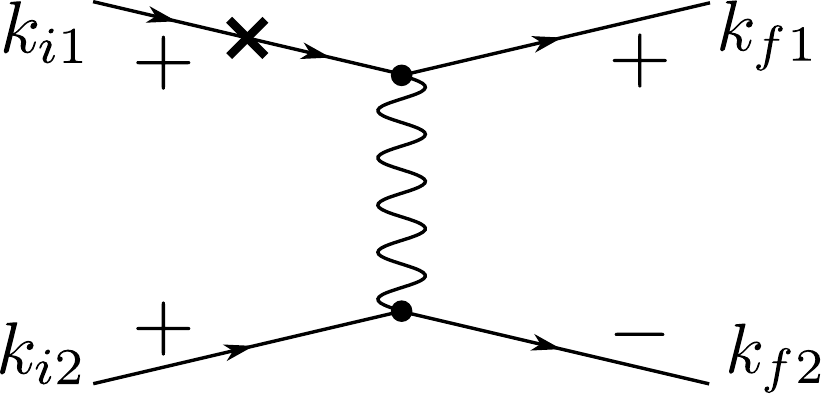}
\caption{Representative one-particle backscattering process to second order in the electron-electron interaction (wavy line) and impurity scattering (cross).}
  \label{fig:HelicalBS}
\end{figure}

In the limit $v_F|k_F|\gg T$, combined processes involving potential scattering off an impurity and scattering by the interaction, provide the leading contribution to $\delta G$. The second iteration of the $\hat{T}$-matrix equation (\ref{tmatrix}) yields a number of cross-terms in $H_V$ and $H_{\tint}$. A typical contribution is shown in Fig.~\ref{fig:HelicalBS}: one incoming particle is scattered at the impurity into a virtual intermediate state, and subsequently interacts with a second incoming particle. Summing all such contributions, we find the effective Hamiltonian $H^\eff_{V,\tint}$ and (for $v_F|k_F|\gg T$)
\begin{align}
 \delta G_{V,\tint} \approx& 1.21 \times 8^{4} \frac{e^2}{h}L n_{\rm imp}\left(\frac{V_0 U_0}{v_F^2}\right)^2 \left(\frac{k_F}{k_0}\right)^{8} \left(\frac{T}{v_F k_0}\right)^4 \label{eq:deltaGintV}
\end{align}
in line with the power-counting argument given after Eq.~(\ref{Heff2}) above. In deriving Eq.~(\ref{eq:deltaGintV}) we assumed that the impurities are randomly positioned along the edge with dilute linear density $n_{\rm imp}$, such that multiple scattering can be neglected. Specifically, interference terms are small for $T\gg v_F/L$ \cite{aronov87}. A comparison of Eqs.~(\ref{eq:deltaGint2}) and (\ref{eq:deltaGintV}) shows that for sufficiently weak impurity scattering, $(n_{\rm imp}/k_0) (V_0/v_F)^2 (T/v_F k_0)^7 \ll 10^{-5}$, the dependence of $\delta G$ on $k_F$ at fixed $T$ displays a minimum at some finite value of $|k_F|$.

The perturbation theory in $U_0$ and $V_0$ diverges if the initial or final state of one of the electrons is at the TR invariant momentum $k=0$, and the intermediate state in Fig.~\ref{fig:HelicalBS} approaches it. This divergence is similar to the one in the cotunneling amplitude in Coulomb blockaded quantum dots, and may be treated in a similar way (see, e.g., Appendix C in Ref.~\cite{koch04}). The interactions lead to a finite lifetime of the intermediate particle or hole at $k=0$ and cut off the divergence. If that lifetime is longer than the time of flight $L/v_F$, then the singularity is cut off by the deviation of the eigenstates from plane waves due to impurity scattering. In any case, higher-order terms in $V_0$ and $U_0$ regularize the divergent contributions and make them smaller than the results in Eqs.~(\ref{eq:Gint})-(\ref{eq:deltaGintV}).

So far, we have used an effective one-dimensional model of the helical edge. However, the edge states exist at the boundaries of 2D topological insulators. Thus, their wave functions decay exponentially on a momentum-dependent length scale $1/\lambda(k)$ into the bulk of the topological insulator, e.g., $\phi_k(x,y) \propto e^{-\lambda(k) y} e^{i k x}$. As a consequence, the spin-rotation matrices $B_k$ involve convolutions over $y$ of the two-dimensional eigenstates along with the 2D interaction potential $U(x,y)$ and impurity potential $V(x,y)$. Importantly, however, the small-$k$ behavior of $B_k$ is always compatible with the expansion (\ref{eq:Bk_approx}), since the latter follows from TR invariance.

In order to justify our effective 1D model (\ref{eq:H0gen})-(\ref{eq:HV}), we determined the momentum-dependent rotation of the spin quantization axis explicitly for the BHZ model \cite{bernevig06} in the presence of Rashba SOI. This model provides a description of 2D topological insulators realized in HgTe/CdTe quantum wells. We used exact diagonalization to solve the Hamiltonian on a cylinder of width $W$, with periodic boundary conditions in the $x$ direction and edges at $y=0$ and $y=W$. In the following, we shall outline the procedure and the results; details will be published elsewhere.

The 2D topological insulators realized in HgTe/CdTe quantum wells can be modeled using a Hamiltonian $H = \sum_k H(k)$, where $H(k)$ has a $4\times4$ matrix structure in a basis containing two spin-degenerate orbitals, $\mathbb{V} = \{ E_\uparrow, H_\uparrow, E_\downarrow, H_\downarrow\}$ \cite{bernevig06}. In the original BHZ model, $H(k)$ is block-diagonal and does not couple spin-up and spin-down states. However, it was shown in Ref.~\cite{rothe10} that breaking inversion symmetry by applying an out-of-plane electric field leads to Rashba SOI and thus a coupling between spin-up and spin-down states. As long as Rashba SOI remains weak, the gapless edge states remain intact. For a given momentum $k$, $H(k)$ has four eigenvectors which correspond to left-moving and right-moving states localized either on the upper or lower edge. These eigenmodes are four-component spinors in the basis $\mathbb{V}$ and will be labeled $\phi_{\alpha,\zeta,k}(x,y)$, where $\alpha = \pm$ denotes the chirality as before, and $\zeta = U,L$ labels the upper or lower edge, at $y=W$ and $y=0$, respectively.

\begin{figure}[t]
  \centering
  \includegraphics[width=8cm]{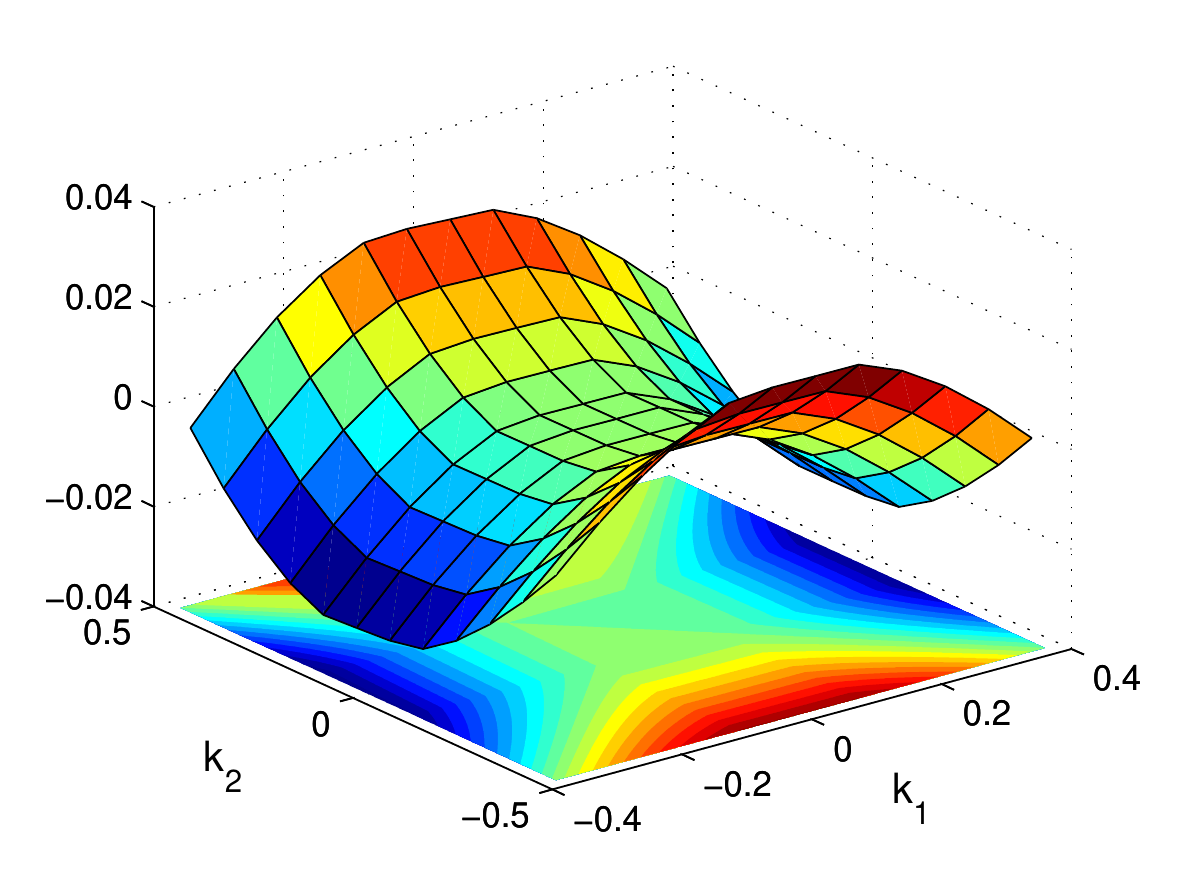}
\caption{(Color online) Off-diagonal component $[B^\dag_{k_1} B_{k_2}]^{-+}$ of the spin-rotation matrix determined from the numerical solution of the BHZ model with Rashba SOI using Eq.~(\ref{eq:B_match}).}
  \label{fig:B_match}
\end{figure}

In order to translate the numerical solution of the 2D model into parameters of the effective 1D model, we match the impurity scattering operator (\ref{eq:HV}) with a corresponding operator in the 2D system. The 1D description of the edge state scattering can be applied if the penetration depth $1/\lambda$ of the edge state into the bulk \cite{qi11} is small compared to the range $l_V$ of the impurity potential. Since we are not interested in the detailed shape of the impurity potential, we assume that the impurity potential is approximately constant over the length scale $1/\lambda$, and that $l_V$ is small compared to the Fermi wavelength in the edge direction. In that case, the potential can be treated as constant in $y$ direction and pointlike in $x$ direction. For an impurity located at position $(x_0,y_0)$ near the upper edge, the 2D scattering operator is then given by $H_{bs} = V_0 \delta(\hat{x} - x_0) \text{diag}(1,1,1,1)$, where $\hat{x}$ is the position operator along the edge. By comparing matrix elements of $H_{bs}$ between the eigenstates $\phi_{\alpha,U,k}(x,y)$ with the matrix elements of $H_V$, we can identify
\begin{align}\label{eq:B_match}
 [B^\dag_{k_1} B_{k_2}]^{\alpha_1\alpha_2} = \int dy \phi^\dag_{\alpha_1,U,k_1}(x_0,y) \phi_{\alpha_2,U,k_2}(x_0,y).
\end{align}
The result of the numerical solution is shown in Fig.~\ref{fig:B_match}. Importantly, it shows the quadratic dependence on $k_1$ and $k_2$, in agreement with the low-momentum expansion (\ref{eq:Bk_approx}).

In addition, we have numerically calculated the momentum-dependent rotation of the spin axis for various other 2D topological insulators with broken $S_z$ symmetry, e.g., the Kane-Mele model in the presence of Rashba SOI, the BHZ model with bulk inversion asymmetry, and the model by \citet{shitade09} for monolayers of sodium iridate. We found similar results as in Fig.~\ref{fig:B_match} in all cases. For small momenta, $B_k$ follows directly from TR invariance and translational invariance, so we expect it to be universally applicable for helical liquids at low energies and weak disorder. As a consequence, the scaling $\delta G \propto T^4$ in Eq.~(\ref{eq:deltaGintV}) can be expected to hold generically for helical liquids with broken $S_z$ symmetry, even if the spin rotation is created by other mechanisms.

We thank I.~Garate and M.~Goldstein for discussions. This work was supported by the NSF DMR Grant No.~0906498. TLS acknowledges support by the Swiss National Science Foundation. SR acknowledges support from the DFG Grant No.~RA 1949/1-1.

\bibliography{paper}

\end{document}